# Improving Belief Propagation Decoding of Polar Codes Using Scattered EXIT Charts

A. Elkelesh*†, M. Ebada*†, S. Cammerer*, S. ten Brink*
*Institute of Telecommunications, Pfaffenwaldring 47, University of Stuttgart, 70569 Stuttgart, Germany
Email: {elkelesh,ebada,cammerer,tenbrink}@inue.uni-stuttgart.de
† These authors contributed equally to this work.

*Abstract*—For finite length polar codes, channel polarization leaves a significant number of channels not fully polarized. Adding a Cyclic Redundancy Check (CRC) to better protect information on the semi-polarized channels has already been successfully applied in the literature, and is straightforward to be used in combination with Successive Cancellation List (SCL) decoding. Belief Propagation (BP) decoding, however, offers more potential for exploiting parallelism in hardware implementation, and thus, we focus our attention on improving the BP decoder. Specifically, similar to the CRC strategy in the SCL-case, we use a short-length "auxiliary" LDPC code together with the polar code to provide a significant improvement in terms of BER. We present the novel concept of "scattered" EXIT charts to design such auxiliary LDPC codes, and achieve net coding gains (i.e. for the same total rate) of $0.4$dB at BER of $10^{-5}$ compared to the conventional BP decoder.

## I. INTRODUCTION

Polar codes, as devised by Arıkan [1], are based on the concept of channel polarization. They have become an active area of research over the past few years owing to the fact that polar codes are the first theoretically proven type of channel codes known to achieve the capacity of an arbitrary Symmetric Binary Input Discrete Memoryless Channel (BI-DMC) under Successive Cancellation (SC) decoding at affordable complexity [1], assuming infinite length codes.

Tal and Vardy [2] introduced a Successive Cancellation List (SCL) decoding algorithm which achieves a significant gain when compared to the conventional SC decoder for finite length codes. With the aid of a high rate Cyclic Redundancy Check (CRC) code they were able to outperform the Maximum Likelihood (ML) decoder of the pure polar code. Several variants of the SCL decoder were proposed to reduce the computational complexity [3], [4], [5]. However, the potential for parallelism of SC-based decoding is limited owing to its inherently sequential decoding manner. As an alternative, Arıkan proposed an iterative decoding algorithm with high potential for parallelism based on Belief Propagation, along the lines of Gallager's Belief Propagation (BP) decoding algorithm [6], [7]. The BP decoder appears to be more suitable for hardware implementation; however, its performance falls behind the SCL-CRC decoder. A lot of effort has been spent on enhancing the performance of finite length polar codes by concatenation with some types of outer codes. In [8], [9], an outer Reed-Solomon code was used with an inner polar code. In [10], an outer LDPC code was used together with an inner polar code to protect *all* of the bit channels; in contrast, [11] uses an LDPC code to only protect the semi-polarized bit channels, which should benefit most from the added redundancy.

In this paper, we propose a novel EXIT chart-based design method to enhance the overall performance of the scheme in [11]. So called "scattered" EXIT charts provide good guidelines for degree profile design even for the *short-length* LDPC codes used. With scattered EXIT charts, the polar/variable node behavior can be predicted, and the check node degree profile can be optimized to match the behavior of the polar/VND, respectively. The proposed design leads to a net coding gain of 0.4dB over the conventional BP decoding of polar codes, and a gain of 0.2dB over [11], all benchmarked within the same simulation framework.

## II. POLAR CODES

### A. Channel Polarization

Channel polarization is a method of constructing $N$ polarized channels out of $N$ identical independent copies of a BI-DMC. The channel polarization theorem states that, as the number of channels tends to infinity, the symmetric capacity of the synthesized channels approaches either $0$ or $1$. The good channels (symmetric capacity close to $1$) or noiseless channels are used to transmit uncoded information bits while the bad channels (symmetric capacity close to $0$) or noisy channels are used to transmit "frozen" (i.e., perfectly known) bits. The frozen bits are a group of non-information bits, and it was proven in [1] that for symmetric channels the value chosen for the frozen vector does not affect the code performance. For simplicity and without loss of generality, the all-zero vector is typically used as the frozen bit vector [1].

### B. Code Construction

Polar code construction is the step of selecting the best $K$ synthesized channels in terms of BER or block error rate (BLER) out of the total $N$ synthesized channels, so that uncoded information bits can be transmitted over these channels, and frozen bits are transmitted over the worst $N - K$ synthesized channels. Therefore, the target of the polar code construction step is to directly minimize the error probability by effectively choosing the channels that will carry the information bits [1], [12].

## C. Polar Encoding

A polar code of length $N = 2^n$ is encoded using the polar code generator matrix $\mathbf{G}$ of size $N \times N$. Thus a block of length $N$, consisting of $N - K$ frozen and $K$ nonfrozen bits, is multiplied by $\mathbf{G}$ to produce the polar codeword. The matrix $\mathbf{G}$ is based on the kernel used to construct the code, with $\mathbf{G} = \mathbf{F}^{\otimes n}$ where $\mathbf{F}^{\otimes n}$ denotes the $n^{th}$ Kronecker power of $\mathbf{F}$, and where $\mathbf{F} = \begin{bmatrix} 1 & 0 \\ 1 & 1 \end{bmatrix}$ is based on the used kernel [1].

## D. Belief Propagation Decoder of Polar Codes

It has been shown, e.g., in [2], [13], that for finite length codes, polar codes are not competitive under SC decoding compared to state-of-the-art LDPC codes. This was the motivation behind proposing different decoding algorithms to enhance the performance of polar codes for finite length codewords. Arıkan suggested a Belief Propagation decoding scheme which is based on Gallager's Belief Propagation (BP) decoding algorithm. Tal and Vardy introduced list decoding of polar codes which was a breakthrough in the polar coding field, making them a strong competitor to LDPC codes [13]. The performance of polar codes under list decoding is better than that under belief propagation decoding [13]. Owing to the sequential decoding strategy of the list decoder, belief propagation still attracts quite some attention, as it offers potential for cost-effective highly parallelized hardware implementation. Also, BP decoding is of potential importance in applications requiring soft-output decoding.

BP decoding of polar codes is a message passing algorithm based on the factor graph of the polar code shown in Fig. 3a, and can be used to estimate the codeword $\hat{\mathbf{x}}$ or the message $\hat{\mathbf{u}}$. For a polar code of length $N = 2^n$, there are $n+1$ stages and $N$ nodes per stage. Messages are iteratively passed between adjacent nodes from left to right and from right to left until a maximum number of iterations is reached; a hard decision is used to estimate the codeword or the message. The factor graph consists of Processing Elements (PEs), each connecting 4 nodes in 2 consecutive stages as shown in Fig. 1 in [14]. The soft messages are updated at each PE as follows:

$L_{out,1} = g(L_{in,1}, L_{in,2} + R_{in,2}); L_{out,2} = g(R_{in,1}, L_{in,1}) + L_{in,2}$
$R_{out,1} = g(R_{in,1}, L_{in,2} + R_{in,2}); R_{out,2} = g(R_{in,1}, L_{in,1}) + R_{in,2}$

where $g(L_1, L_2) = \ln\left(\frac{1 + e^{L_1 + L_2}}{e^{L_1} + e^{L_2}}\right)$ is commonly referred to as "box-plus" operator. For the $g(\cdot)$-function a min-sum approximation $g(L_1, L_2) = \text{sign}(L_1) \cdot \text{sign}(L_2) \cdot \min(|L_1|, |L_2|)$ can be used which is more suitable for hardware implementation [14].

The BP decoder propagates soft messages between the adjacent nodes, with two types of messages involved: left-to-right messages, called $\mathbf{R}$-messages, and right-to-left messages, called $\mathbf{L}$-messages. The $\mathbf{L}$-messages in the Log-Likelihood Ratio (LLR) domain of the $N$ nodes in stage $n+1$ are the channel output LLR messages given by $L_{n+1,j} = \ln\left(\frac{P(y_j|x_j=0)}{P(y_j|x_j=1)}\right)$ for $1 \leq j \leq N$, where $P(y_j|x_j)$ denotes the conditional probability that the channel output $y_j$ is received when the codeword element $x_j$ was transmitted. The values of the $\mathbf{R}$-messages of the $N$ nodes at stage 1 are either zero or infinity for the nonfrozen or frozen bit channels, respectively.

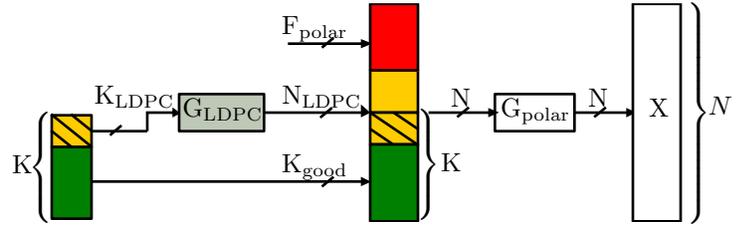

Figure 1: Polar encoder with auxiliary LDPC code.

The decoding starts by propagating the $\mathbf{R}$-messages from left to right, and then propagating the $\mathbf{L}$-messages from right to left and so on, until reaching a maximum number of iterations; then a hard decision is applied to the estimate of the codeword or the message bits given as follows

$$\begin{aligned} L(\hat{u}_i) &= L_{i,1} + R_{i,1} \\ L(\hat{x}_i) &= L_{i,n+1} + R_{i,n+1} \end{aligned} \quad (1)$$

where $L(\hat{u}_i)$ and $L(\hat{x}_i)$ are the LLRs of the estimated message and the estimated transmitted codeword, respectively [14], [15], [16].

## III. APPLYING AN AUXILIARY LDPC CODE OVER SEMI-POLARIZED CHANNELS

A lot of attention has been given to the problem of reducing the gap between the performance of the BP decoder and the SCL decoder. Due to the sequential hierarchical structure of the SCL decoder, appending a high rate auxiliary CRC code is an elegant way for quickly checking whether the correct codeword is among the proposed list of codewords, resulting in a significant improvement in terms of BER performance (or, equivalently, in terms of net coding gain). However, a seamless integration of the CRC code into the BP decoder, which would require a trellis structure with a large state space, appears to be much less possible. In fact, for the BP decoder, an auxiliary code with, again, a graph-based BP decoding strategy (e.g., an LDPC code) is a much better match.

To enhance the BP decoder performance in a similar fashion akin to the CRC code improving SCL decoding, polar codes were concatenated with different types of codes. In [11], a short LDPC code is proposed to protect only those bits transmitted on the semi-polarized channels, leading to a 0.3dB gain compared to the conventional BP decoder. In fact, the appended LDPC code is not an outer code in the sense that an outer code handles the whole information bits prior to the "inner" polar code; rather, one would consider it as a short *auxiliary* code that improves protection only of the semi-polarized channels.

The polarization of bit channels becomes better when $N$ becomes large. However, for short-to-moderate length polar codes, some channels are fully polarized (either noiseless, or very noisy channels) while others remain in an intermediate,

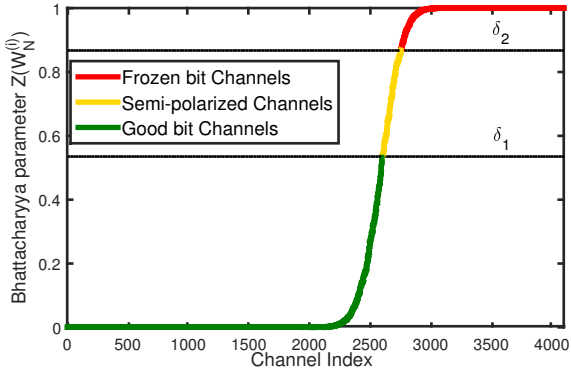

Figure 2: Sorted Bhattacharyya parameters $Z\left(W_N^{(i)}\right)$ for $N = 4096$.

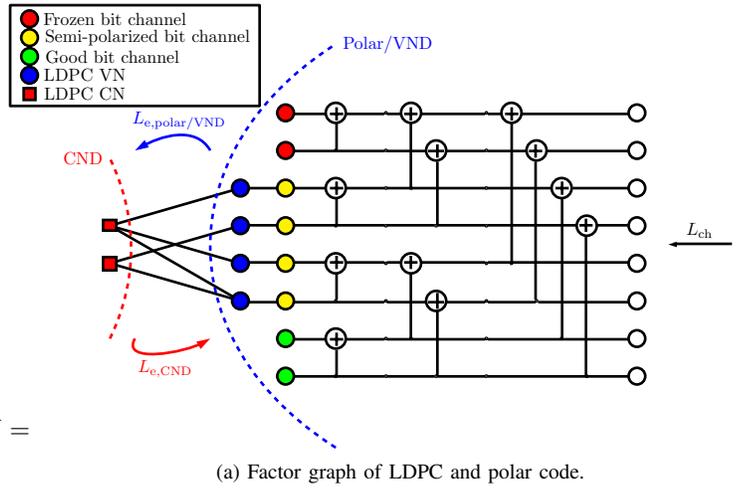

(a) Factor graph of LDPC and polar code.

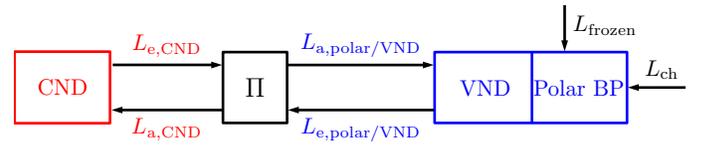

(b) Information flow in the iterative decoder.

Figure 3: BP information flow of the combined polar and LDPC decoder.

"semi-polarized" stage. When information is transmitted over these intermediate channels, the probability of a bit error is still rather high. The set-up in [11] is based on protecting these intermediate channels by appending a short-length LDPC code. This leaves potential for further optimizing the appended LDPC code, to improve the overall BP decoder performance.

The encoder structure is shown in Fig. 1. As usual, uncoded bits are transmitted on the good channels ($K_{good}$), while the known, frozen bits (zeros here) are transmitted on the weak channels ($F_{polar}$). A short-length auxiliary LDPC code protects the information bits transmitted over the semi-polarized channels. Thus, $K_{LDPC}$ bits are encoded using an LDPC encoder to produce $N_{LDPC}$ bits to be conveyed on the semi-polarized channels.

The semi-polarized channels, upon which LDPC coding is applied, are the channels with intermediate Bhattacharyya parameter value $Z\left(W_N^{(i)}\right)$ as shown in Fig. 2 following the basic set-up in [11]. For a specific set of thresholds $\delta_1$ and $\delta_2$, with $0 < \delta_1 \leq \delta_2 < 1$, three sets of channels can be defined:

1) good channels, $Z\left(W_N^{(i)}\right) \leq \delta_1$
2) intermediate channels, $\delta_1 < Z\left(W_N^{(i)}\right) \leq \delta_2$, and
3) bad channels, $Z\left(W_N^{(i)}\right) > \delta_2$.

The total encoding rate is given by $R_{total} = \frac{K_{good}+K_{LDPC}}{N}$ and, throughout this paper, fixed to $1/2$; the LDPC code is of rate $R_{LDPC} = \frac{K_{LDPC}}{N_{LDPC}}$, and the polar code of rate $R_{polar} = \frac{K_{good}+N_{LDPC}}{N}$.

The corresponding decoder is shown in Fig. 3a and is just an extended version of the normal polar decoding factor graph. The BP decoder (or "Tanner graph") of the auxiliary LDPC code is connected to the leftmost stage of the BP polar decoder. First, the **R**-messages propagate from left to right, then the **L**-messages propagate from right to left until reaching stage 1; then LLR messages $L_{i,1}$ are passed on to the LDPC decoder to perform one LDPC BP decoder iteration. The extrinsic information of the LDPC BP decoder is fed back to stage 1 ($R_{i,1}$) of the polar BP decoder as shown in Fig. 3b. One polar BP iteration is followed by one LDPC BP iteration until reaching a maximum number of iterations; finally, a hard decision is taken to estimate the codeword $\hat{\mathbf{x}}$ or the message $\hat{\mathbf{u}}$ according to (1), respectively.

## IV. SCATTERED EXIT CHARTS

An Extrinsic Information Transfer (EXIT) chart [17] is an efficient tool to design concatenated codes based on the analysis of their iterative decoding behavior [18]. EXIT charts track the exchange of extrinsic information between two or more decoders, and allow to predict whether the overall decoding process converges. The decoding is successful if there exists an open tunnel between both EXIT curves [17]; code optimization is possible by matching the EXIT curves of the individual component decoders. A decoding trajectory obtained by simulating the actual iterative decoder verifies whether the prediction holds true.

EXIT designs are based on asymptotic behavior, and thus long codewords are one prerequisite for "classic" EXIT charts to be meaningful. Due to the short-length of the auxiliary LDPC code, the decoding trajectories will vary a lot, and deviate significantly from the predicted (asymptotic) behavior. Furthermore, the "inner" polar/VND-curve of the set-up of Fig. 3b cannot be easily estimated because of internal states of the polar BP decoder, namely $L_{i,j}$ and $R_{i,j}$, which make it hard to obtain the extrinsic output simply based on a priori and channel information as inputs, respectively; i.e., the polar BP decoder has *memory* built up from prior uses at earlier stages of the iterative decoding process. Contrary to this, aside

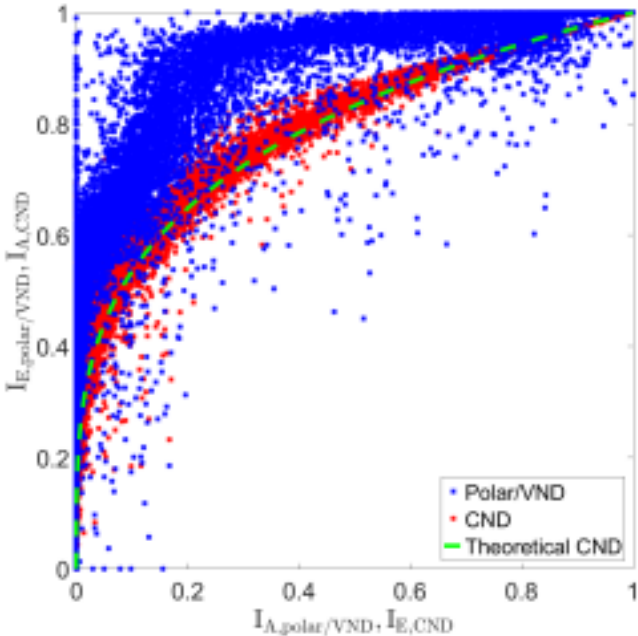

Figure 4: Blue: Polar/VND ($d_v = 3$) scattered EXIT curve; red: CND ($d_c = 5$) scattered EXIT-curve; green: theoretical CND-curve ($d_c = 5$); $N_{LDPC} = 155$, $E_b/N_0 = 2.5$dB.

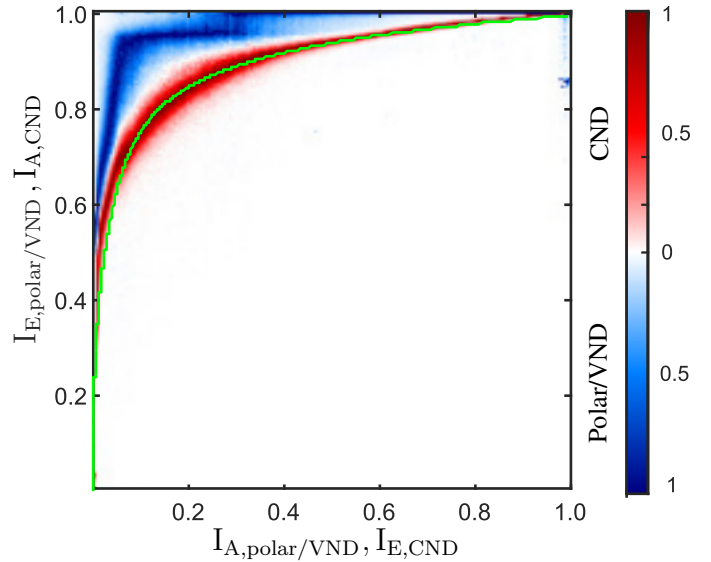

Figure 5: Frequency of occurrence (2D-histogram) for CND- and polar/VND scattered EXIT-curves of the proposed check-irregular code; green: CND EXIT curve-with $\overline{d_c} = 10.72$

from short-length effects, the CND EXIT-curve is well known and analytical models have been derived in the literature. To mitigate both issues (short-length effects and internal memory of the polar BP decoder) while still obtaining meaningful EXIT predictions, we resort to the novel concept of "scattered" EXIT charts. A scattered EXIT chart (Fig. 4) uses the statistics of numerous EXIT trajectories obtained from simulations of the actual iterative decoder, and tracks their frequency of occurrence in a two-dimensional histogram over the EXIT mutual information plane.

While the individual trajectories would resemble an erratic, chaotic behavior (not shown) which is not useful for code design, the statistical averaging effects traced by the scattered EXIT chart allow us to "see through the clutter" and obtain a good estimate of the polar/VND-EXIT behavior, as depicted in Fig. 5. Using that estimate, one can find a better matching CND-curve, and thus, in the same spirit as classic EXIT charts, improve the overall performance of the iterative decoder. Note that Fig. 5 indicates how well the analytical description of the CND-curve coincides, on average, with the simulated CND behavior, even for short codeword length $N_{LDPC}$.

## V. NUMERICAL RESULTS

Throughout this paper, we use the polar code construction based on Arıkan's Bhattacharyya bounds [1] of bit channels designed at $E_s/N_0 = 0$dB, a polar codeword length of $N = 4096$, and an overall rate of $1/2$. The EXIT chart acts as a guide through the optimization process by matching the CND-curve to the polar/VND-curve obtained from the scattered EXIT chart.

### A. Obtaining the polar/VND-curve

The scattered EXIT chart is used to obtain the polar/VND-curve, and thus to select *that* CND-curve (i.e., check node degree, or degree profile) which matches best. The goal is to reduce the "unused" area in the open convergence tunnel. As depicted in Fig. 4, the scattered EXIT CND-curve of the LDPC code ($d_c = 5$) agrees well with the expected theoretical CND-curve. One can now optimize the whole set-up based on the, implicitly obtained, polar/VND-curve. Although the variance of the scattered EXIT polar/VND-curve seems large, it clearly varies around a deterministic curve, as can be seen from the two-dimensional histogram of Fig. 5.

### B. Matching the CND-curve to the Polar/VND-curve

As we are unable to predict the polar/VND-curve analytically, we resort to focusing on the CND-part. Closing the open tunnel between both curves requires increasing the check node degree, which, on the other hand, leads to an intersection of both curves at high BER regions, causing the convergence to fail. This motivated us to use an *irregular* check node decoder: Including some lower check node degrees keeps the tunnel close to the polar/VND-curve without causing an intersection, as can be inferred from in Fig. 6.

### C. Comparing BER performance

Tab. I shows the simulation parameters as described in Fig. 1. The first set-up is used for the polar code BP-only simulation. The second set-up uses the same code parameters as in [11], which is a polar code augmented by a 155 bit regular $(3,5)$ LDPC code. The third set-up is our new design: a polar code augmented by a 155 bit ($d_v = 3, \overline{d_c} = 10.72$)-irregular LDPC code (a combination of check node degrees 4, 5 and 17 according to 0.3322, 0.1628 and 0.505 of all check nodes). Finally, the fourth set-up is a further improved version of the third set-up, now using a 190 bit LDPC code instead of 155 bits.

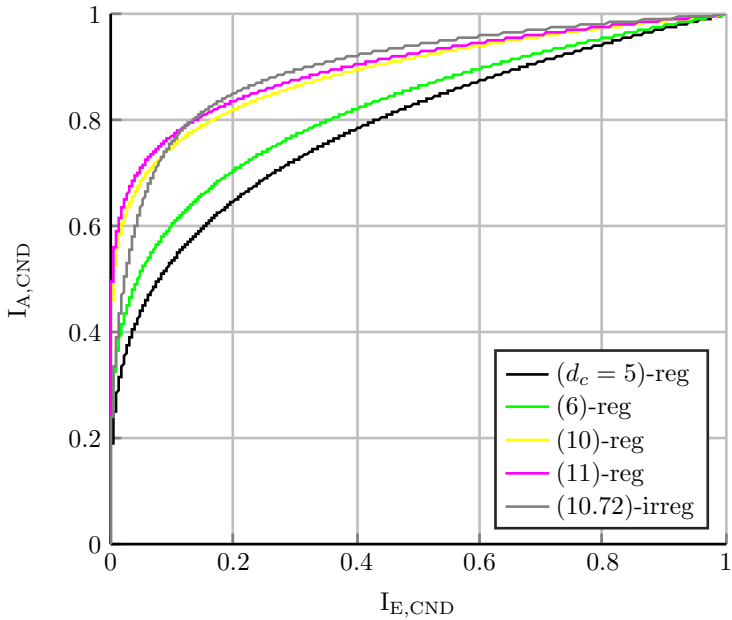

Figure 6: Expected CND-curves for different check node degrees (and profiles).

Table I: Simulation parameters used for different set-ups.

| set-up | $K_{LDPC}$ | $N_{LDPC}$ | $(d_v, \overline{d_c})$ | $K_{good}$ | $F_{polar}$ |
|---|---|---|---|---|---|
| 1 | 0 | 0 | n/a | 2048 | 2048 |
| 2 | 62 | 155 | $(3,5)$-reg | 1984 | 1957 |
| 3 | 112 | 155 | $(3,10.72)$-irreg | 1934 | 2007 |
| 4 | 137 | 190 | $(3,10.72)$-irreg | 1910 | 1996 |

From the scattered EXIT chart in Fig. 4 we obtained that a $(d_v = 3, \overline{d_c} = 10.72)$-irregular LDPC code of length 155 appended to the polar code should work well (CND-curve, see Fig. 5). And indeed, as can be seen from Fig. 7, an additional net coding gain of $0.2$dB at BER of $10^{-5}$ can be achieved compared to the second set-up as used in [11], which, by itself, already has a $0.3$dB gain over the conventional BP decoder (set-up 1). Increasing $N_{LDPC}$ (set-up 4) further enhances the overall system performance; an additional gain can be obtained when $N_{LDPC}$ is slightly increased.

## VI. CONCLUSION

We introduced the novel concept of scattered EXIT charts and used it for refining the check node degree profile of an auxiliary LDPC code as appended to a polar code. Regardless of using short-length LDPC codes, we showed that scattered EXIT charts are useful for optimizing the overall set-up, even without having an analytical polar/VND-curve available. A check-irregular LDPC code was designed, leading to a net coding gain of $0.4$dB at a BER of $10^{-5}$ compared to the conventional BP decoder.

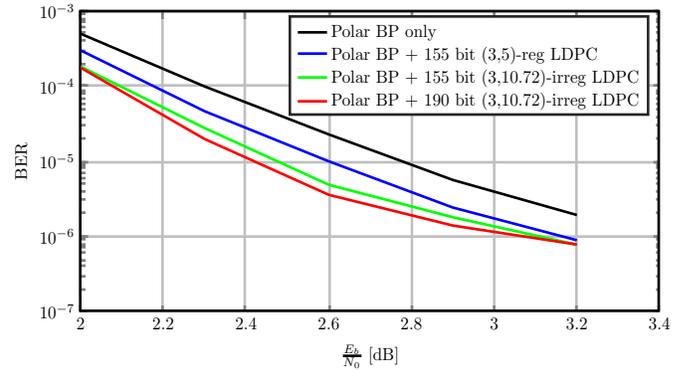

Figure 7: Simulated BER of the different set-ups (compare to Tab. I).